\documentclass[showpacs,amsmath,amssymb,preprint,aps,showkeys]{revtex4}
\usepackage{graphicx}
\usepackage{dcolumn}
\usepackage{bm}

\begin{document}


\title{Effects of a cosmological constant on the X-ray fluorescence spectra of black hole accretion disks}
\author{Sergei Slobodov}
\email{slobodov@physics.ubc.ca}
\author{Kristin Schleich}
\email{schleich@noether.physics.ubc.ca}
\author{Donald M. Witt}
\email{donwitt@noether.physics.ubc.ca}
\affiliation{Department of Physics and Astronomy, University of British Columbia, Vancouver, British Columbia, Canada V6T 1Z1}

\date{\today}

\begin{abstract}
We investigate the effects of cosmological constant on the characteristic peaks in the iron line profile in the black hole accretion disk X-ray spectrum. Our results show that for a fixed mass black hole, the peaks become less pronounced and closer together with increasing cosmological constant. This effect is mainly due to the slower rotational velocity of Keplerian orbits at large radii in the Schwarzschild-de Sitter spacetime as compared to those for the Schwarzschild spacetime. This change of the iron line profile is similar to that obtained from increasing the outer radius of the accretion disk size or reducing the emission intensity power law exponent in the  accretion disk emissivity models. 
\end{abstract}


\maketitle

\section{\label{sec:intro}Introduction}

Observing the iron $K_\alpha$ line in X-ray fluorescence emitted from the accretion disk around a black hole is an attractive way to probe the geometry of spacetime in this region, because this normally narrow line is seen by a remote observer as a characteristic two-peak profile flourescence spectrum \cite{Fabian:1989ej},\cite{Tanaka:1995en}. This profile carries the imprint of the spacetime metric in the vicinity of the emitting region; the curvature of spacetime affects this line both by Doppler broadening from the moving scatterer in the accretion disk and by gravitational redshift of the scattered photons. Quantitative calculations of this iron line profile have been previously obtained for two natural choices of black hole backgrounds: the Schwarzschild and Kerr solutions (see e.g. \cite{Fabian:1989ej}, \cite{Fanton}, \cite{CHF}, \cite{Laor:1991}). The resulting spectra are qualitatively the same, but are potentially observationally distinguishable with high resolution spectral observations. In recent years, with the launch of Chandra and XMM-Newton, the quality of the observed spectra is beginning to approach the resolution neccessary to discriminate between these two background spacetimes \cite{Fab2000}. 

It is natural to ask whether or not other background spacetimes will produce identifiable features in precision measurements of iron line profiles. In this paper we will derive the formalism needed for the calculation of the spectrum of a narrow emission line in a static spherically symmetric metric. We then apply this formalism to consider the effects of the cosmological constant $\Lambda$ on the line shape of the accretion disk spectra. In section \ref{sec:basics} we present the necessary  formalism and in section \ref{sec:SdS} we apply these results to the case of the Schwarzschild-de Sitter metric. In section \ref{sec:spectrum} we descibe our numerical algorithm, an adaptation of \cite{CE} for numerically computing the spectra. In section \ref{sec:results} we present our calculated line profiles. We find that the cosmological constant modifies the line profile for values of the dimensionless parameter $\Lambda M^2 > 10^{-7}$ that characterizes this effect.  At the present time, WMAP observations give a value of $\Omega_\Lambda= .73 \pm .04$ \cite{Bennett:2003bz}. This value corresponds to $\Lambda = 1.3\times 10^{-46} km^{-2}$ in units where $G=c=1$.  Thus this cosmological constant  causes deviations from the Schwarzschild line profile only for a black hole that is unrealistically massive, $10^{19}$ solar masses. Alternately, for black holes in galactic cores with masses of $10^6$ to $10^8$ solar masses, a cosmological constant $23$ or more orders of magnitude greater than the current WMAP value is needed to cause such deviations. Although these results clearly imply that iron line spectra will not provide an independent measurement of the cosmological constant, they do point out that metric deviations from other spherically symmetric sources could produce a measurable effect.

\section{\label{sec:basics}The Iron Line Spectrum}
To calculate the observed X-ray florescence spectrum from an accretion disk, we need to add the flux reaching the remote observer from every infinitesimal element of the disk. The intensity and the frequency of flux received from each infinitesimal element is determined by a variety of factors: the rotational velocity of the emitter, its direction relative to the observer, the lensing and the gravitational red shift effects of the black hole, and the emissivity of disk at the position of that infinitesimal element. Following e.g. \cite{Fabian:1989ej}, we assume that the fluorescing particles  travel along the circular timelike geodesics (Keplerian orbits), that the disk itself is angularly uniform, stationary and thin, and that the wavelength of the emitted photons is short enough that geometric optics applies (See \cite{Fabian:1989ej} for the discussion of these assumptions.). There are a variety of approaches to explicitly carrying out this calculation: \cite{Fabian:1989ej}, \cite{Fanton}, \cite{CHF}, \cite{CE}, \cite{Cunn}. We follow that of Chen and Eardley,\cite{CE}, with minor modifications as we find it intuitive and easily generalized to an arbitrary static spherically symmetric metric.

The frequency shift $g$ of a photon travelling from emitter to observer is given by
\begin{equation}
\label{eq:redshift}
g = \frac{\nu}{\nu_{e}} =\frac{-p \cdot u_{o}}{-p \cdot u_{e}},
\end{equation} 
where $\nu_e$ is the frequency of the photon in the emitter's frame, $\nu$ is the frequency of the photon as measured by the observer,  $p$  is the 4-momentum of the photon along its null geodesic and $u_{obs}$ and $u_{em}$ are the 4-velocities of the observer and the emitter respectively.

The total bolometric flux $F$ seen by an observer is linked to that emitted from the source by a factor of $g^4$ due to conservation of 4-volume \cite{MTW}. Hence, the observed  and emitted  intensities at a given frequency are related to each other by a factor of $g^3$.

The observed line flux at a given frequency from the accretion disk can be readily expressed in terms of the specific intensity $I(\Omega,r,\nu_e)$ in the emitter's frame and the frequency shift $g$:

\begin{equation}
 F_\nu = \int g^3 I(\Omega,r,\nu_e) d\Omega,
\end{equation}
where the integration is over the angles subtended by the disk image. Note that $F_\nu$ will depend on the inclination of the disc relative to the line of sight of the observer. 

\begin{figure}
\centering
\includegraphics{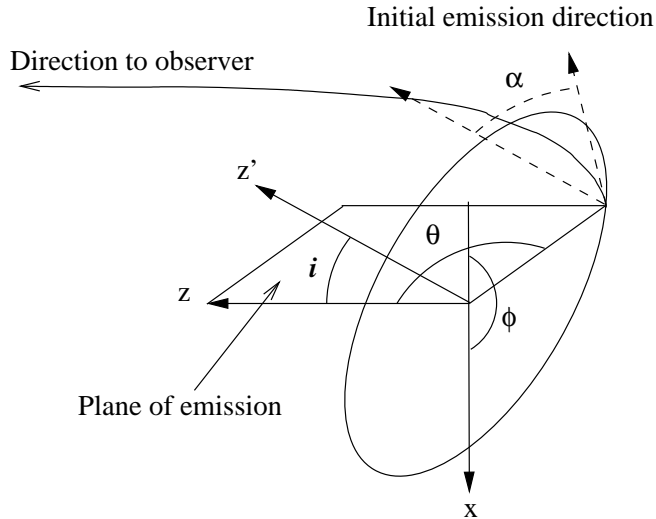}
\caption{Emission angle $\alpha$ and inclination angle $i$ in the spherical coordinate system of the observer (\ref{eq:metric}). Here $z$ is the observer's $z$-axis, $z'$ is the accretion disk's axis, $x$ is the $x$-axis in either coordinates and $i$ is the disk inclination.\label{fig:coord}}
\end{figure}

To proceed further, explicit expressions for the intensity and frequency shift are necessary; we will derive this first for a general static, spherically symmetric metric. These calculations can be considerably simplified by a convenient choice of the coordinates. Following \cite{CE}, we use a spherical coordinate system $(t,r,\theta,\phi)$ centered around the gravitational source and $z$-axis pointing towards the observer. Further, without loss of generality, we choose the accretion disk axis to be in the $y=0$ plane, with the axis projection towards negative $x$. (See Fig.~\ref{fig:coord}, based on Fig.1 of \cite{CE}.)
The inclination angle $i$ is the angle between $z$-axis and the disk axis. 

The general static spherically symmetric metric can be written in these coordinates as
\begin{equation}
\label{eq:metric}
ds^2=-T(r)dt^2+R(r)^{-1}dr^2+r^2d\Omega^2.
\end{equation}
Note that spherical symmetry implies that all geodesics will be planar. 
The 4-velocity of a circular timelike geodesic in the accretion disk 
at the radial coordinate $r$ in this metric can be expressed using the linear rotational velocity of a circular timelike emitter $v$ 
\begin{equation}
\label{eq:v}
v=\sqrt{\frac{rT'}2}
\end{equation}
and a factor $\gamma$
\begin{equation}
\label{eq:gamma}
\gamma=\frac{1}{\sqrt{T-v^2}}
\end{equation}
as
\begin{equation}
\label{eq:em4vel}
u^a_{e} = \gamma 
(t^a +\frac 1r v\cos\alpha \theta^a +\frac 1{r\sin\theta}v\sin\alpha \phi^a),
\end{equation} 
where $t^a$,$r^a$, $\theta^a$ and $\phi^a$ are the coordinate basis vectors and $\alpha$ is the angle between the plane of the disk and the plane defined by the emitting point, the center of the disk and the observer. It can be expressed in terms of $i$ and $\phi$ as
\begin{equation}
\label{eq:alpha}
\cos\alpha = \sin{i}\sin\phi.
\end{equation} 
The polar angle of the intersection in these coordinates is given by
\begin{equation}
\cot{\theta} = \cos{\phi}\tan{i};
\end{equation}

The 4-momentum of a photon emitted from the accretion disk is
\begin{equation}
\label{eq:ph4mom}
p^a = h\nu(\frac 1T t^a + \sqrt{\frac RT - R(\frac br)^2}r^a +\frac{b}{r^2}\theta^a),
\end{equation} 
where $b$ is the photon impact parameter. The corresponding null geodesic lies in the plane connecting the observer, the emitter and the center of the disk.

There are two natural choices for the 4-velocity $u_{o}$ of a timelike observer in a spherically symmetric metric: that corresponding to an observer maintaining a fixed distance $d$ from the black hole,
\begin{equation}
\label{eq:sobs4vel}
u^a_o= \frac 1{\sqrt{T(d)}} t^a
\end{equation}
and that corresponding to a radially freely falling observer at distance $d$,
\begin{equation}
\label{eq:obs4vel}
u^a_{o} = \frac{\epsilon}{T(d)}t^a \pm \sqrt{R(\frac{\epsilon^2}{T(d)}-1)}\ r^a
\end{equation} 
where $\epsilon$ is a constant determined by the initial radial velocity. Note that if the spherically symmetric metric is not asymptotically flat, (\ref{eq:sobs4vel}) will not approach the timelike Killing vector $t^a$ for large $d$. 
These two 4-velocities coincide for the choice of zero  radial velocity at $d$, $\epsilon = \sqrt{T(d)}$. In the following we will use (\ref{eq:obs4vel}).

The frequency shift is conveniently written in terms of two factors:
\begin{equation}
g = g_{obs}g_{disk}, 
\end{equation} 
where for a given photon impact parameter $b$, $g_{obs}$ depends on $d$ and $\epsilon$ only, 
\begin{equation}
\label{eq:gglobal}
g_{obs} = \frac{\epsilon}{T(d)}-\sqrt{\frac 1{T(d)}(\frac{\epsilon^2}{T(d)}-1)}
\sqrt{1-T(d)\frac{b^2}{d^2}}
\end{equation} 
and the expression for $g_{disk}$ depends on the emitter's position $r$ only:
\begin{equation}
\label{eq:glocal}
g_{disk} = \frac{1}{\gamma(1-\frac{b}{r}v\sin{i}\sin\phi)}.
\end{equation}

We will assume that the iron line emission in the emitter's frame is monochromatic and isotropic with no local line broadening. For this case the dependence on radius $r$ of the specific intensity is commonly modelled by a power law dependence. Assuming an optically thin disk, 
\begin{equation}
\label{eq:in}
I(\Omega,r,\nu_e) = \frac \kappa{4\pi\mu'}r^{-p}\delta(\nu_e-\nu_0).
\end{equation}
The power $p$ is a real number usually taken to be in the range of $1\le p \le 3$; $\kappa$ is a constant with units of power. 
The factor $\mu'$ is the cosine of the angle between the emitted photon and the disk axis in the emitter's frame,
\begin{equation}
\mu'=\frac {gb\cos i}{r \sin\theta}.
\end{equation}
An explicit derivation of the form $\mu'$ in the observer's coordinates is given in Ref. \cite{CHF}.

The infinitesimal solid angle subtended by the disk is easily expressed in terms of the impact parameter $b$ and $\phi$:
 \begin{equation}
d\Omega =\frac{ bdbd\phi}{d^2}
\end{equation}
Combining terms and reexpressing $\nu_e = \nu/g$ in (\ref{eq:in}) we have
\begin{equation}
\label{eq:diskflux}
 F_\nu = \frac \kappa{4\pi d^2}\int g^3 \frac {r \sin\theta}{b\cos i} r^{-p}\delta(\nu-g\nu_0) bdbd\phi \ 
\end{equation}
using the identity $\delta(\nu/g -\nu_0) = g \delta (\nu - g\nu_0)$. The range of integration is over $\phi$ and $b$ such that the null geodesics intersect the disk.
The coordinates $r,\theta$ of the intersection point are implicitly a
function of $b$, $\phi$, and $i$; a null geodesic of given $b$ and $\phi$ will intersect the accretion disc at angle $i$ at a given $r$ and $\theta$. For a given frequency $v$, contributions to $F_v$ are from points on the accretion disk with equal value  of $g$ (See \cite{Luminet} Fig.7 for an example of such contours for Schwarzschild.). Observe that the intersection of the null geodesic with the accretion disc can occur either before or after the turning point.

\section{\label{sec:SdS}The Schwarzschild - de Sitter Metric}
We now apply the results of section \ref{sec:basics} to the Schwarzschild - de Sitter metric. This metric describes a static spherically symmetric vacuum solution of Einstein equations with cosmological constant. Cosmologically, this is a good model of an isolated black hole in a dark energy-dominated spatially flat universe. The Schwarzschild - de Sitter metric in static coordinates is
\begin{equation}
\label{eq:SdSmetric}
ds^2=(1-\frac{2M}{r}-\frac{1}{3}\Lambda r^2)dt^2+\frac{dr^2}{1-\frac{2M}{r}-\frac{1}{3}\Lambda r^2}+r^2d\Omega^2\ .
\end{equation}

The effective potential for this spherically-symmetric metric is 
\begin{equation}
\label{eq:Veff}
V_{eff}=\frac{l^2}{2r^2}-\frac{M}{r}-\frac{M l^2}{r^3}-\frac{1}{6}\Lambda{r^2}
\end{equation}
where $l$ is the conserved orbital angular momentum. Circular orbits correspond to the extrema of this effective potential $V_{eff}$:
\begin{equation}
\frac{M}{r^2}-\frac{\Lambda r^2}{3}-\frac{l^2}{r^3}+\frac{3Ml^2}{r^4}=0\ .
\end{equation} 
The circular orbit is stable when the extremum is a minimum:
\begin{equation}
\label{eq:stab}
\frac{\Lambda}{3}(4-\frac{15M}{r})<\frac{M}{r^3}(1-\frac{6M}{r})\ . 
\end{equation} 
For $\Lambda M^2=0$,  we recover the familiar circular orbit stability limit for the Schwarzschild metric: $r>6M$ \cite{MTW}. We now examine in more detail the orbit stability for the case where the spacetime is close to that of Schwarzschild, namely that of small cosmological constant relative to the black hole mass, $2M << \sqrt{\Lambda/3}$ and distances far from the black hole, $d>>2M$ \footnote{There is a limit on how big $\Lambda$ can be for an accretion disk to appear at all. For example, for large $\Lambda$ the term $(1-\frac{2M}{r}-\frac{1}{3}\Lambda r^2)$ is negative 
for all r, thus making the t coordinate spacelike.}. See, for example, \cite{Stuchlik:1999qk} for a detailed analysis of geodesics in the Schwarzschild-de Sitter spacetime. 

Unlike in the case of the Schwarzschild solution, circular orbits become unstable at both small and large radii in the metric. The instablility at small radii occurs as for the Schwarzshild solution. For small enough $\Lambda M^2$, the minimum stable circular orbit  is approximately $r_{min} \simeq 6M + 648 \Lambda M^3$. In contrast to the Schwarzshild solution, there is a new instablilty at large radii; the ``negative pressure'' produced by the cosmological constant destabilizes the outer circular orbits.  From (\ref{eq:stab}) we can calculate that the maximum possible stable orbit is approximately at
\begin{equation}
\label{eq:outerstab}
r_{max}=(\frac{3M}{4\Lambda})^ \frac{1}{3}
\end{equation} 
Objects beyond this limit would either find a smaller stable circular orbit or become unbound. This effect limits the value of $\Lambda$ for which the Schwarzchild-de Sitter metric makes sense as a background for a black hole with an accretion disk. For black holes of between  $1$ and $10^{10}$ solar masses, a range from stellar remnant black holes to the mass of active galactic nuclei, the current estimated value of $\Lambda$ yields an $r_{max}$ of the order of $20$ to $40,000$ parsec. 

The linear rotational velocity  seen by the observer of an element of the disk at radius $r$ is 
\begin{equation}
\label{eq:linearvel}
v= \sqrt{\frac{M}{r}-\frac{\Lambda r^2}{3}},
\end{equation} 
or slower than the velocity at the same radius in the Schwarzschild metric, $v=\sqrt{\frac{M}{r}}$. This is as expected, as the cosmological term acts as a repulsive force and partially screens the central mass \footnote{At $r=({\frac{3M}{\Lambda}})^\frac{1}{3}$ the central mass is completely screened, so that a particle can "hover" at that distance without rotating at all. This distance, however, is farther from the center than the maximum stable circular orbit, and so the equilibrium is unstable.}.

To calculate the spectrum, we need to choose an observer. We will assume that the observer has zero radial velocity relative to the black hole; with this assumption, the observer's 4-velocity coincides with (\ref{eq:sobs4vel}) and $g_{obs}$ becomes
\begin{equation}
g_{obs}=\frac 1{\sqrt{1-\frac {2M}d - \frac {\Lambda}{3} d^2}}
\end{equation}
independent of the integration parameters of (\ref{eq:diskflux}).
The contribution from the disk is 
\begin{equation}
\label{eq:glocal-dS}
g_{disk} = \frac{\sqrt{1-\frac{3M}{r}}}{1-\frac{b}{r}\sqrt{\frac{M}{r}-\frac{1}{3}\Lambda r^2}\sin{\alpha}} \ .
\end{equation} 

As $g_{obs}$ is independent of the integration parameters in (\ref{eq:diskflux}) it can be factored out of the integral; (\ref{eq:diskflux}) becomes
\begin{equation}
\label{eq:diskflux2}
 F_\nu = \frac {\kappa g_{obs}^3}{4\pi d^2}\int g_{disk}^3 
 \frac {r \sin\theta}{b\cos i} r^{-p}\delta(\nu-g\nu_0) bdbd\phi \ .
\end{equation}
The factor $\frac {\kappa g_{obs}^3}{4\pi d^2}$ depends on both the distance to the accretion disk $d$ and $\Lambda$; the integrand is now independent of $d$.
\section{\label{sec:spectrum}The Numerical Spectrum Calculation Procedure}

Each emission point of the accretion disk "seen" by a remote observer corresponds to a distinct pair of impact parameter and polar angle $(b, \phi)$. Thus sampling the $(b, \phi)$ space is equivalent to sampling the whole disk. This approach, due to \cite{CE}, circumvents the need to explicitly evaluate the Jacobian of coordinate transformation from the disk coordinates to observer coordinates. In other approaches (\cite{Fabian:1989ej}, \cite{Fanton}, \cite{Cunn}) this Jacobian is calculated either numerically or analytically.
\begin{figure}
\centering
\includegraphics{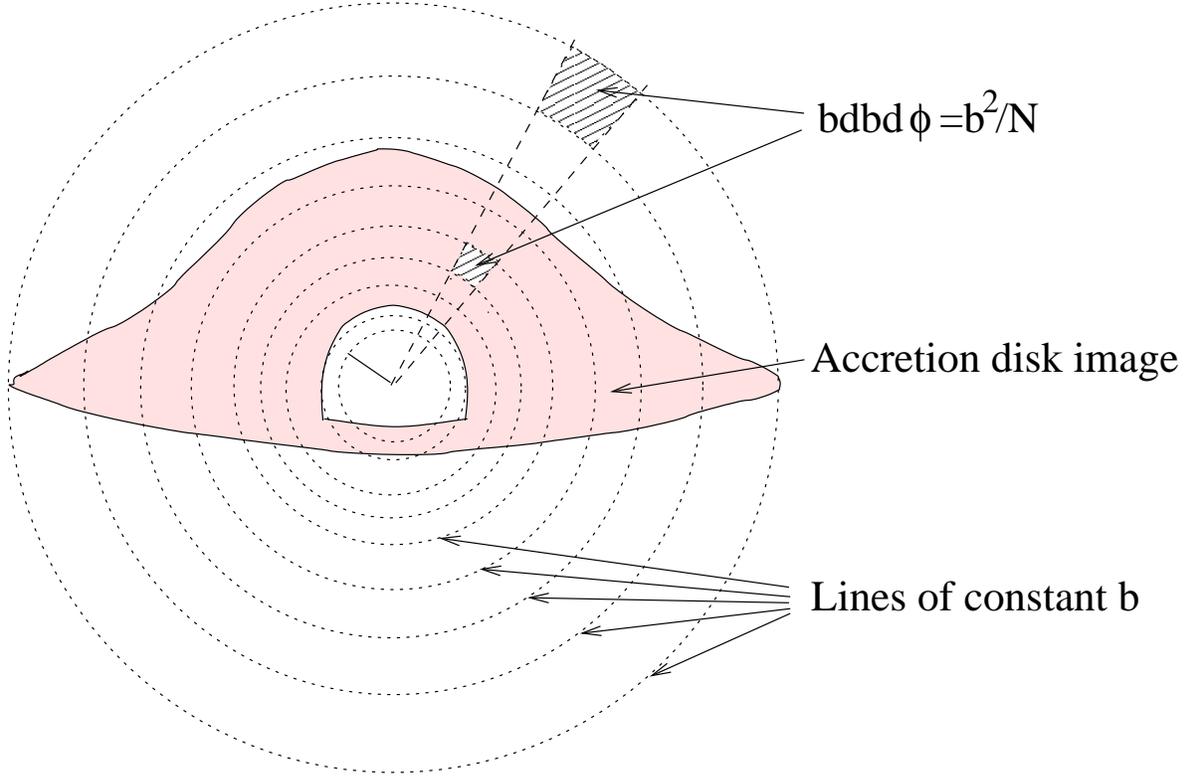}
\caption{For numerical calculations the accretion disk image is divided into $N^2$ nearly square elements whose boundaries are lines of constant $b$ and constant $phi$. The area of each element is chosen to be proportional to the square of $b$.\label{fig:bdbdphi}}
\end{figure}

The spectrum calculation algorithm is as follows:
\begin{enumerate}
\item Specify the metric and the disk parameters: the desired number of disk image elements to use, the inner and outer disk radii in dimensionless units, $r_{in}, r_{out}$,
the inclination angle $i$ of the disk and the emission intensity power law exponent $p$ (see equation (\ref{eq:in})).
\item Sample the $(b, \phi)$ space. To accomplish this, we discretize $d\phi =  2\pi/N$ and take $db=2\pi~b/N$. The impact parameter $b$ ranges from zero to a value determined by the outer disk radius, chosen so that no null geodesic with greater $b$ will intersect the disk. The flux contribution from each element and its frequency shift is taken to be that of center of each element. 
\item For each sample point, integrate the null geodesic inward from the observer to to the emission point, defined as the point where the geodesic intersects the disk. (Points in $(b, \phi)$ space with null geodesics found not to intersect the disk at this step are discarded.) 
\item Calculate the frequency shift using  equation (\ref{eq:glocal-dS}) and the spectral flux from the disk element using equation (\ref{eq:diskflux}). Add the resulting intensity to the appropriate frequency point of the spectrum.
\end{enumerate}

The usual way of numerically integrating the null geodesic is to use the second-order geodesic equations in order to avoid complications near turning points, especially for the high-deflection angle case (see, for example, \cite{CE}, \cite{MTW}, \cite{Luminet}). In our case, however, nearly all of the flux comes along null geodesics with small to moderate deflection, because the inner radius of the accretion disk is at least at $6M$. Thus, the null geodesic
equation
\begin{equation} 
\label{eq:nullODE}
\frac{dr}{d\theta}= {\pm}b\sqrt{(\frac r b)^2+\frac{2M}{r}-1+\frac{\Lambda~r^2}3},
\end{equation} 
can be integrated by first-order methods. To produce a "bounce" from a turning point, we change the sign of $\frac{dr}{d\theta}$ from $-$ to $+$ when the expression under the square root in \ref{eq:nullODE} becomes negative. This approach is stable under changes in stepsize and reproduces certain analytic deflection results to high accuracy.

The code implementing this algorithm is available from the authors. 

Note that we modeled the disk image as a collection of small nearly-square elements in the space $(b,\phi)$ (see Fig.~\ref{fig:bdbdphi}), and calculated the flux from each one, assuming uniform flux from each piece. The error introduced by this procedure adds noise to the resulting line profile. Higher noise level for the same discretization is seen for large outer radii and for high inclination angle. This is because, for a given number of disk elements, fewer image elements overlap the important inner disk area for a large disk. Similarly, for a large inclination angle, fewer elements overlap the disk, causing a higher noise level. However, we observed no measurable systematic error even for significant discretization noise. Our solution was to increase the number of points (and line profile computation time) from the usual choice of $10^6$ to $10^8$ in such cases. 

\begin{figure}
\includegraphics{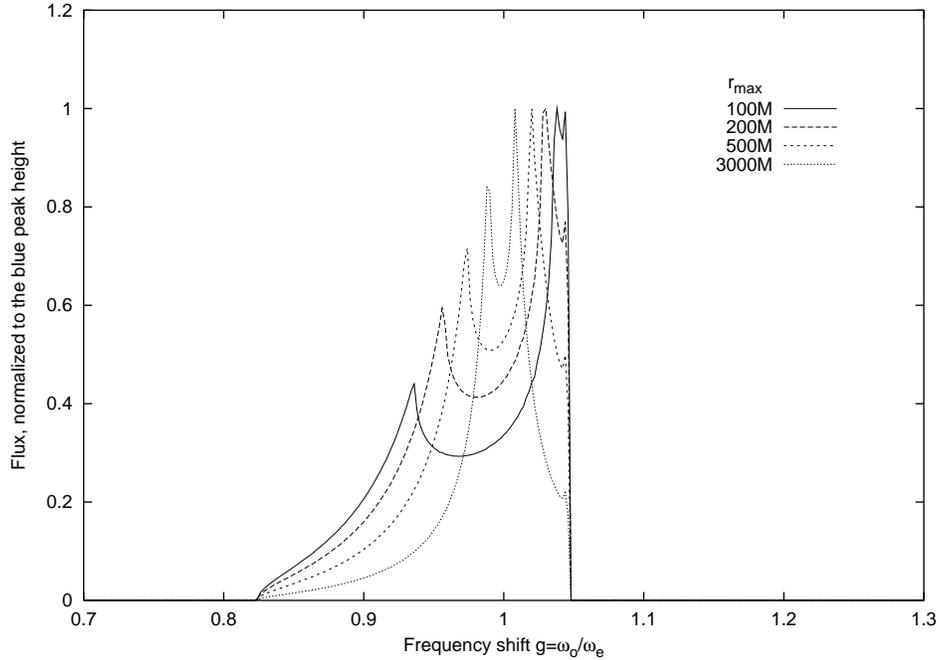}%
\caption{Spectra calculated for the Schwarzchild metric with parameters taken from 
\cite{Fabian:1989ej} Fig.1b: $i=30^o$, $p=2$, $r_{min}=20M$, $r_{max}=100M, 200M, 500M, 
3000M$ (left to right).\label{fig:fabian-1b}}
\end{figure}

For algorithm verification purposes, we calculated line profiles for the Schwarzschild metric with the parameter sets taken from \cite{Fabian:1989ej} Fig.1, \cite{CE} Fig.4a-7a \footnote{The captions of \cite{CE} Fig.4 and Fig.6 appear to be swapped in print.}. Our simulation results for the Schwarzschild case are indistinguishable from those in the original papers. For example, a set of line profile spectra calculated using the parameters taken from  Fig.1b of \cite{Fabian:1989ej} is shown on Fig.~\ref{fig:fabian-1b} and replicates the spectra from that paper. We feel confident that our implementation of the accretion disk fluorescence spectrum calculations is solid enough to apply it to the Schwarzschild - de Sitter case.

\section{\label{sec:results} Results for the Schwarzschild - de Sitter Metric}
In the Schwarzschild case, the line profile of the black hole normalized to the height and position of the blue peak, does not depend on the mass $M$ of the black hole. Instead, it depends on the three dimensionless parameters: the ratios of inner and outer accretion disk radius to the black hole mass and the emissivity parameter $p$. The emissivity parameter $p$ determines the sensitivity of the line profile to the other disk radii. The line profile for  $p=2$ (\cite{Fabian:1989ej}) is more sensitive to the choice of inner and outer radii than that for $p=3$ (\cite{CE}); this is easily understood as the total flux from a fixed radius falls off as $r^{2-p}$.

The line profile in the Schwarzschild-de Sitter metric is similar to that of
the Schwarzschild case; however the normalized line profile now depends on four dimensionless parameters. The additional parameter is $\Lambda M^2$ as discussed in section \ref{sec:spectrum}. In addition, note that the absolute position of the blue peak in frequency will in principle depend on both the distance to the observer and the cosmological constant. However, for the range of $\Lambda M^2$ in our simulations, this effect is negligible.
 
\begin{figure}
\includegraphics{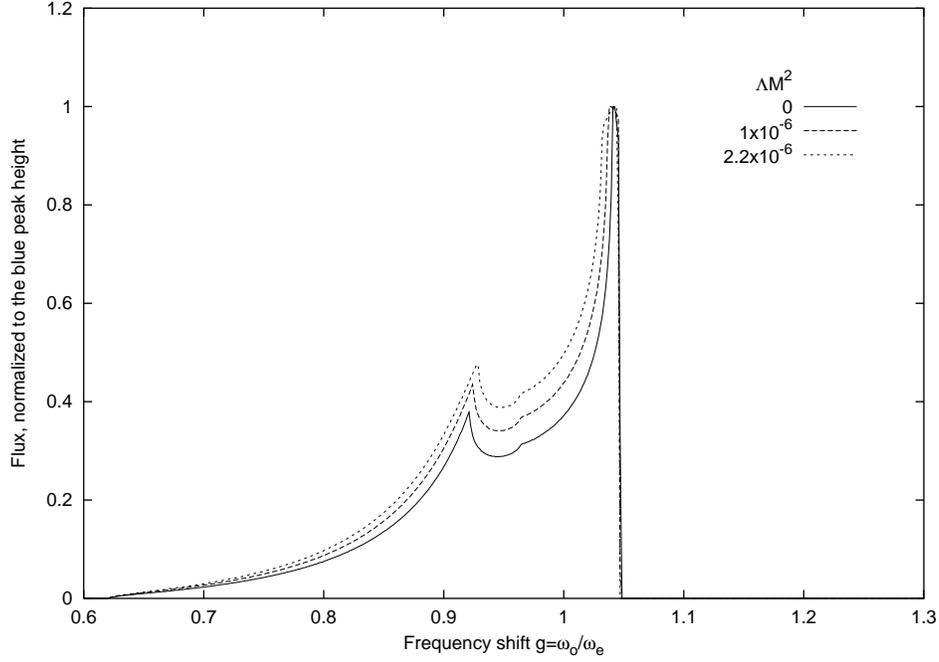}%
\caption{Spectra calculated for a range of $\Lambda{M}^2$ for inclination angle $i=30^o$, $p=2$, $r_{min}=7M$, $r_{max}=70M$. Note that the line for $\Lambda{M}^2= 1 \times 10^{-7}$ nearly coincides with that for the Schwarzschild case on this and other graphs.\label{fig:lambda-1}}
\end{figure}

Fig.~\ref{fig:lambda-1} shows the effects of cosmological constant on the normalized accretion disk fluorescence spectra for the disk of inner radius $7M$ and outer radius $70M$, fixed disk inclination $i=30^{\hbox{\rm o}}$ and a constant emissivity index $p=2$. The parameter $\Lambda M^2$ is varied from $1\times 10^{-8}$ to  $2.2\times 10^{-6}$. This maximum  value of $\Lambda M^2$ corresponds to an maximum outermost stable orbit of  radius $70M$ (cf. \ref{eq:outerstab})).  This data set is similar to that in Fig.1 of \cite{Fabian:1989ej}. However, for most of the data sets used in \cite{Fabian:1989ej,CE}, the  outer radius exceeds the radius of the outermost stable circular orbit values of $\Lambda M^2 > 10^{-7}$, which is barely large enough to produce a noticeable difference in the spectrum. We reduced outer disk radius to just $70M$ to allow for larger $\Lambda M^2$ and more pronounced effects from the cosmological constant.
The simulation shows that the effects are negligible for $\Lambda M^2$ below about $10^{-7}$. Above that, the relative height of the red peaks get progressively larger with the increase of the cosmological term. In addition, the peaks tend to get closer together. 

\begin{figure}
\includegraphics{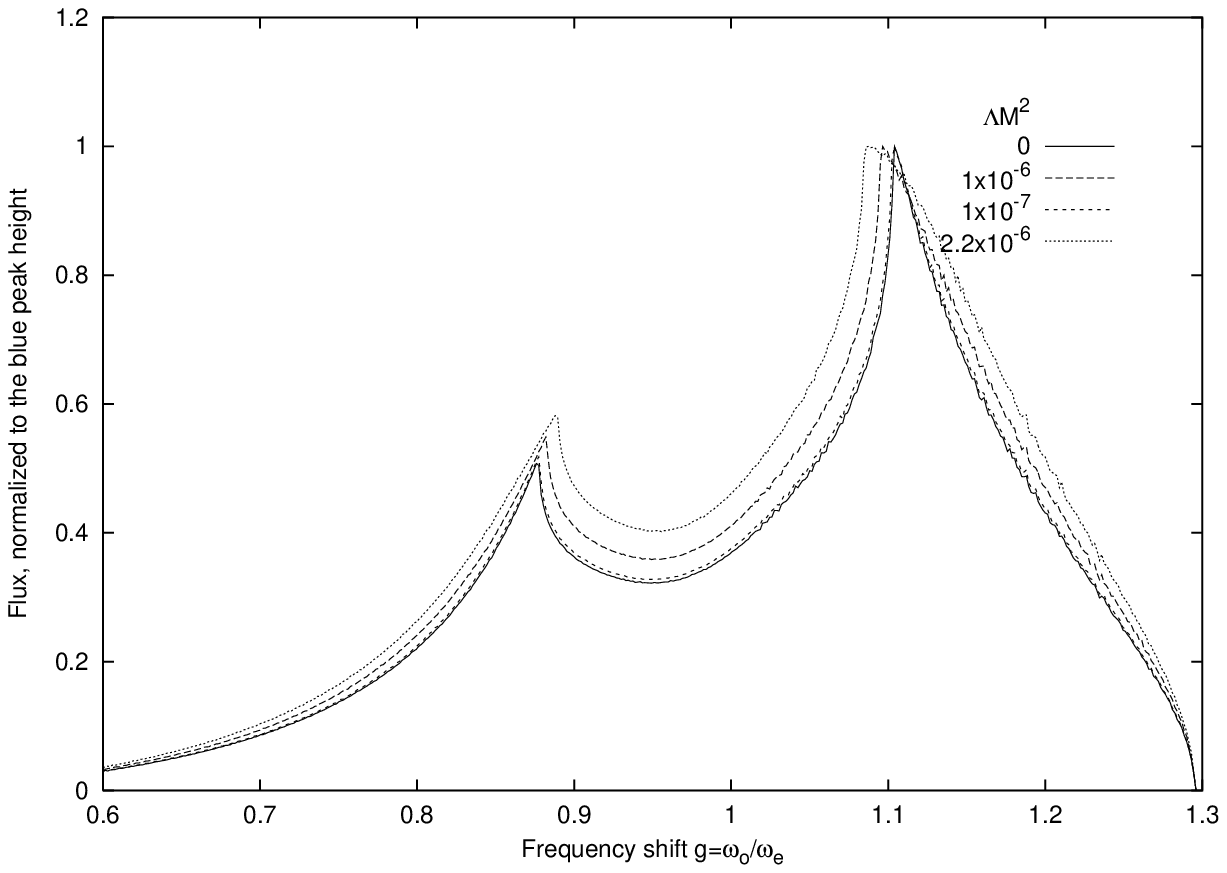}%
\caption{Spectra calculated for a range of $\Lambda{M}^2$ for inclination angle  $i=70^o$, $p=2$, $r_{min}=7M$, $r_{max}=70M$.\label{fig:lambda-2}}
\end{figure}

\begin{figure}
\includegraphics{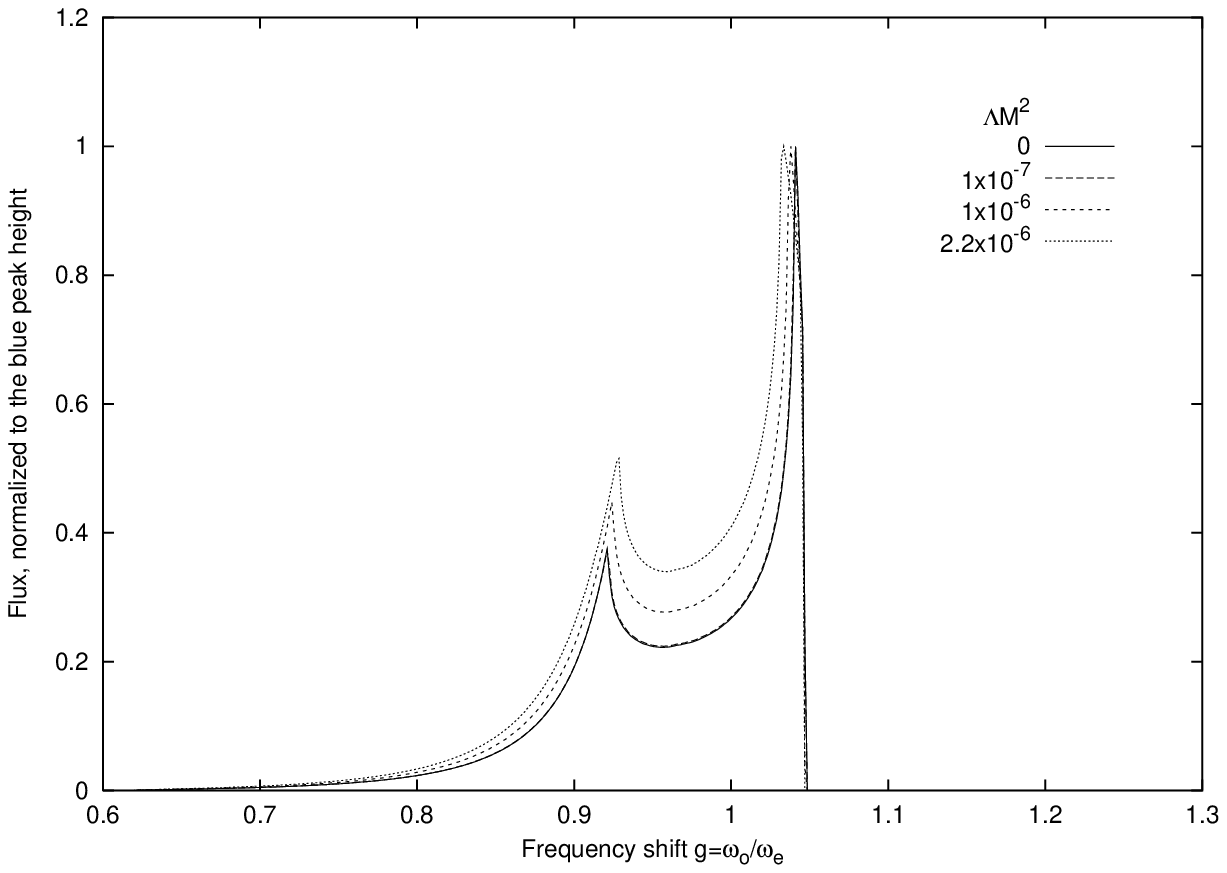}%
\caption{Spectra calculated for a range of $\Lambda{M}^2$ for inclination angle $i=30^o$, $p=1$, $r_{min}=7M$, $r_{max}=70M$.\label{fig:lambda-3}}
\end{figure}

\begin{figure}
\includegraphics{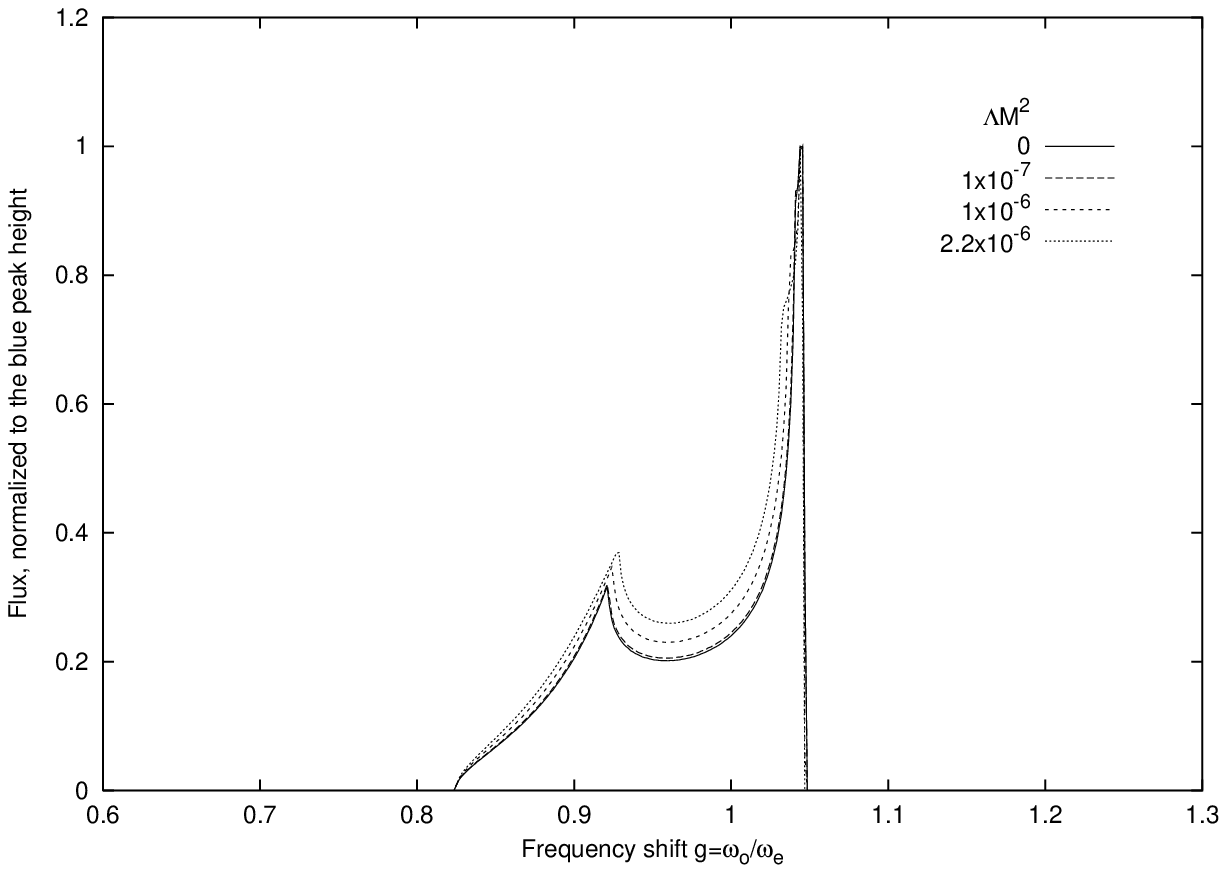}%
\caption{Spectra calculated for a range of $\Lambda{M}^2$ for inclination angle $i=30^o$, $p=2$, $r_{min}=20M$, $r_{max}=70M$.\label{fig:lambda-4}}
\end{figure}

Figures~\ref{fig:lambda-2}-\ref{fig:lambda-4} show the dependence of spectra on the cosmological term for  the specified values of  inclination angles, emissivity index and  disk dimensions. The results are similar to that illustrated in equation (\ref{eq:linearvel}) for all cases. 

\begin{figure}
\includegraphics{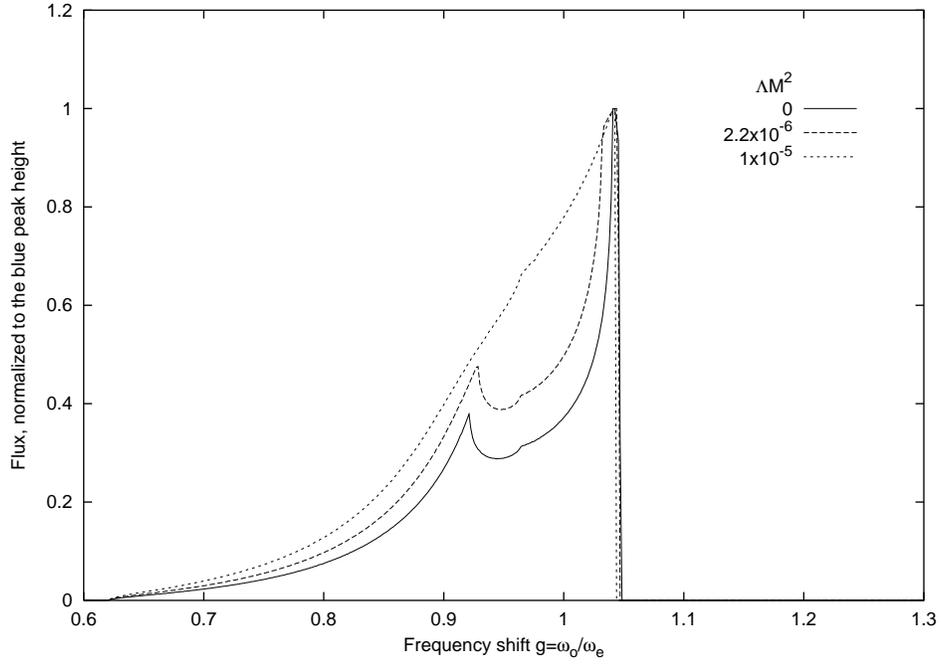}%
\caption{Spectra calculated for $\Lambda{M}^2=10^{-5}$ for inclination angle $i=30^{\circ}$, $p=2$, $r_{min}=7M$, $r_{max}=70M$. Note that the maximum radius exceeds the value allowed by the orbit stability condition for $\Lambda{M}^2=1\times 10^{-5}$.\label{fig:lambda-5}}
\end{figure}

As discussed earlier, the data above are limited to the parameter sets that ensure stable circular orbits for all parts of the disk. Much more dramatic differences from Schwarzschild spectra can be obtained if the condition of Keplerian orbit stability at the outer disk is neglected. Consider, for example, a line profile for $\Lambda M^2=10^{-5}$ and $p=2$ illustrated in Fig. \ref{fig:lambda-5}. This line profile deviates significantly from the familiar two peak line profile seen generally. Of course, it is very likely that extending the accretion disk into region with unstable orbits is unphysical.
\section{\label{sec:discussion}Discussion}
One can intuitively understand the effects of the cosmological constant as follows: $\Lambda$ acts as a negative pressure, reducing  the rotational velocity of the disk at a given radius in comparison to the  Schwarzschild case. The farther away a circular geodesic is from the black hole, the more pronounced is the velocity reduction. Now most of the material is concentrated in the outer layers of the disk and as the red (blue) peak is produced by the parts of the disk moving fastest away (towards) the observer,  slower rotational velocity farther away means less pronounced peaks. This effect is somewhat similar to that of reducing the maximum outer radius or increasing the emissivity index, both of which shift the relative contribution to the spectrum in favor of the inner parts of the disk.

One can estimate the magnitude of the cosmological constant required to see its effect on the iron line profile for black holes. Assuming the cosmological constant value consistent with the current observations, This value corresponds to $\Lambda = 1.3\times 10^{-46} km^{-2}$ in units where $G=c=1$, calculated from the WMAP-based values for the dark energy density $\Omega_{\Lambda}=0.73 \pm 0.04$  and the Hubble constant $h=(0.71 \pm 0.04)100 km/s/Mpc$ \cite{Bennett:2003bz}.  We find that a solar mass black hole, i.e. order of magnitude size of astrophysical black holes with observed iron line profiles yields $\Lambda M^2 = 2.9\times 10^{-46}$, an undetectable effect. For black holes of $10^8$ solar masses, roughly the upper limit of galactic black hole masses,  yields $\Lambda M^2 = 2.9\times 10^{-30}$. To produce a deviation from the Schwarzschild accretion disk iron line profile, this factor has to larger than $10^{-7}$, which requires either an increase in $\Lambda$ by 23 orders of magnitude or an increase in the black hole mass by over 11 orders of magnitude to $3 \times10^{19}$ solar masses. Thus observations of the line shape of the iron line profile will not provide sensitive independent bounds on the cosmological constant.  This conclusion also holds for other homogeneous dark energy models, which, as they have similar dark energy densities (\cite{Kunz:2003iz}), have similar metrics.

Although these results clearly imply that iron line spectra will not provide an independent measurement of the cosmological constant, they do provide an example how spherically symmetric metric deviations can potentially produce a measurable effect. 

\begin{acknowledgments}
The authors would like to thank the Natural Sciences and Engineering Research Council of Canada for its support.
\end{acknowledgments}

\bibliography{SdS-paper-v2}

\end{document}